# Linear Instability of the Plane Couette and Plane Poiseuille Flows

Sergey G. Chefranov [1)], Alexander G. Chefranov [2)]

[1)] A.M. Obukhov Institute of Atmospheric Physics RAS, Moscow, Russia;
e-mail: schefranov@mail.ru
[2)] Eastern Mediterranean University, Famagusta, North Cyprus; e-mail: Alexander.chefranov@emu.edu.tr

September 29, 2015

## Abstract

We show possibility of the Plane Couette (PC) flow instability for Reynolds number $Re > Re_{th} \approx 140$. This new result of the linear hydrodynamic stability theory is obtained on the base of refusal from the traditionally used assumption on longitudinal periodicity of the disturbances along the direction of the fluid flow. We found that earlier existing understanding on the linear stability of this flow for any arbitrary large Reynolds number is directly related with an assumption on the separation of the variables of the spatial variability for the disturbance field and their periodicity in linear theory of stability. By the refusal from the pointed assumptions also for the Plane Poiseuille (PP) flow, we get a new threshold Reynolds value $Re_{th} \approx 1040$ that with 4% accuracy agrees with the experiment contrary to more than 500% discrepancy for the earlier known estimate $Re_{th} \approx 5772$ obtained in the frame of the linear theory but when using the "normal" disturbance form (S. A. Orszag, 1971).

## Introduction

Until now, there exists a problem of understanding the mechanism of losing stability of a laminar flow resulting in the transition to a new turbulent regime of the medium flow. Actually, in the existing up to now linear hydrodynamic stability theory for the Hagen-Poiseuille (HP) flow in a round pipe, transition from a laminar regime to any other flow regime can't happen for any finite threshold Reynolds number $Re_{th}$ that does not comply with the experimental data. Similar conclusion of the linear theory is also available for the Plane Couette (PC) flow [1]. For the Plane Poiseuille (PP) flow, known estimate of the threshold Reynolds number made on the base of the linear theory more than five times greater than the value obtained in the experiment [1-3].

In the result, the pointed transition between the laminar and turbulent regimes for the HP, PC, and PP flows is described up to now only on the base of nonlinear theory of finite amplitude disturbances [1]. However, up to now, it is not proved the very fact that such transition shall always correspond to this hard finite-amplitude mechanism of the stability losing for HP, PC, and PP flows, and not to the soft mechanism described by the linear theory for the disturbances arbitrary small by amplitude.

We show in the present paper that linear instability of the PC flow for the finite Reynolds numbers $Re > Re_{th}$ is possible but only under condition of refusing from the traditional "normal" disturbance form (for which, periodical variability of the disturbance field along the direction of the main flow is assumed together with separation of variables describing spatial disturbance variability). Earlier a similar conclusion for the HP flow was already obtained under this very condition of refusing from the assumption of separation of the spatial variables [4 – 6]. In the frame of the new linear hydrodynamic stability theory, we get a condition of the linear instability of the PP flow that significantly better agrees quantitatively to the experimental data than the previously known one [2].

In the developed here linear theory, we consider an opportunity of the quasi-periodic longitudinal (along the direction of the main flow) variability when the longitudinal and transversal spatial variables are not separated in the disturbance field description. Energy method is used that accounts for existence of various periods of spatial variability for different transversal modes corresponding to evolution of extremely small transversal velocity field disturbances and zero boundary conditions on the solid boundary surfaces for PC and PP flows.

Thus, a fundamental and applied problem of the turbulence emergence mechanism definition for the HP, PP, and PC flows more than a century was remained open due to the stated by the linear hydrodynamic stability theory conclusion contradicting to the experiment both qualitatively (for HP and PC) and quantitatively (for PP).

In the present paper, as well as in [4-6] for the HP flow, we show that for the PC and PP flows, a dissipative instability (DI) mechanism may be the very mechanism of the linear instability for the flows. In its turn, the DI mechanism is typical for many physical systems where an important role is played by disturbances with negative energy [7-9] earlier related with the notion of secular instability [7].

## Problem definition

Let us consider usual representation (see [1]) of the PP and PC flows of viscid incompressible fluid bounded along axis $z$ by two parallel solid surfaces being on distance $H$.

For the PP flow directed along axis $x$, the origin is selected at the middle of the layer (where the flow velocity is maximal), and the static solid boundary surfaces have coordinates $z=H/2$, $z=-H/2$.

For the PC flow, the origin also is selected at the middle of the layer, and we assume that the flow velocity at the layer middle is zero. As in [3], we assume that the solid boundary has the coordinate $z=-H/2$ and moves with the velocity $-V_{max}$, whereas another solid boundary (at $z=H/2$) has velocity $V_{max}$ along the axis $x$.

Linear stability of the flows is investigated in the simplest case when there exist only extremely small disturbances of the transversal velocity field component directed along axis $y$. Let velocity and pressure fields do not depend on that transversal (with respect to the main flow) coordinate $y$. Let us instead of the normal" form consider quasi-periodic disturbances for which transversal and longitude (along the main flow direction) variables are not separated when describing their spatial variability.

In the dimensionless form, we have for the disturbance field the following equation and boundary conditions:

$$\frac{\partial V_y}{\partial \tau} + U(z)\mathrm{Re}\frac{\partial V_y}{\partial x} = \Delta V_y; \Delta = \frac{\partial^2}{\partial z^2} + \frac{\partial^2}{\partial x^2};$$
$$V_y(z=\pm 1) = 0, \qquad (1)$$
$$U(z) = 1 - z^2, \quad for\,PP; U(z) = z, for\,PC.$$

In (1), $V_y$ is non-dimensional velocity obtained by norming the dimensional disturbance velocity by $V_{max}$. The spatial coordinates are un-dimensioned by $H/2$, and the time variable in (1), $\tau = 4t\nu/H^2$, where $t$ is the dimensional time, $\nu$ is the kinematic viscosity coefficient, and $\mathrm{Re} = \frac{V_{max}H}{2\nu}$ is the Reynolds number for the PP and PC flows. In the considered example, there are no other velocity field components. Note, that in [2,3], another transversal disturbance component, not $V_y$, as here, but $V_z$, is considered and that leads to the necessity in [2,3], instead of relatively simple equation (1), to solve significantly more complicated Orr-Sommerfeld equation.

Let us represent the linear equation (1) solution in the following complex form satisfying the boundary conditions on $z$:

$$V_y = e^{\lambda \tau} V; V = \sum_{n=0}^{N}(A_n(x)\sin(\pi z n) + B_n(x)\cos(\frac{\pi z(2n-1)}{2})); \lambda = \lambda_1 + i\lambda_2; V = V_1 + iV_2, V^* = V_1 - iV_2;$$

$$A_n(x) = A_n(x+T_n); B_n(x) = B_n(x+T_n); T_n = 1/\alpha_n; \max T_n = 1/\alpha_0; \alpha_0 < \alpha_1 < ... < \alpha_N; i^2 = -1 \quad (2)$$

In (2) the new periodic boundary conditions are defined on x, which are established and different for each mode with number n.

Thus, instead of the traditional "normal" disturbance representation (with the common for all the modes periodic boundary condition when in (2), $T_n = T = const$ for all $n$), we introduce in (2) $N$ periodic boundary conditions along the axis $x$, that are defined individually for each transversal mode number $n$ ($n=0,1, 2,..., N$).

The new problem definition of the linear hydrodynamic theory problem in the form (1), (2) for the longitudinal quasi-periodic disturbances differs from the ones used earlier and may better agree with the experimental data (where actually quasi-periodic only, not pure periodic, spatial disturbance amplitude variability along the fluid flow direction between two solid planes is observed [1, 3]).

### Energy method

1. Let us consider, on the base of (1), (2), the mean energy (with respect to mass unity) evolution:

$$E = \langle V_y V_y^* \rangle / 2 = e^{2\lambda_1 \tau} \langle VV^* \rangle / 2;$$

$$I_0 = \langle VV^* \rangle = \frac{1}{2}\int_{-1}^{1} dz \frac{1}{T_{\max}} \int_{0}^{T_{\max}} dx VV^* \quad (3)$$

In (3), we can assume that $T_{\max} = \frac{1}{\alpha_0}$.

From (1), for the PP flow, we can get the following equation for exponent $\lambda_1$, defining growth (for $\lambda_1 > 0$) or decay (for $\lambda_1 < 0$) of the energy with time (for simplicity, for PP flow, we can set $B_n = 0$ and sum over $n$ starting from c $n=1$ in all the sums):

$$2\lambda_1 I_0 = I_1 \operatorname{Re} - I_2; I_2 = -\langle V^* \Delta V + V \Delta V^* \rangle > 0$$

$$I_1 = -\langle (U(z) \frac{\partial (VV^*)}{\partial x} \rangle = -\frac{1}{T_{\max}} \sum_{n=1}^{N} \sum_{\substack{m=1 \\ m \neq n}}^{N} q_{nm}(A_n(T_{\max})A_m^*(T_{\max}) - A_n(0)A_m^*(0)), \quad (4)$$

$$q_{nm} = \frac{1}{2}\int_{-1}^{1} dz(1-z^2)\sin(\pi z n)\sin(\pi z m) = \frac{4(-1)^{n+m+1} nm}{\pi^2(n^2 - m^2)^2} \text{ if}: n \neq m;$$

Let us consider for simplicity the case when in (2)–(4), $A_n(x) = A_{0n} \exp(i 2\pi \alpha_n x)$. Then, from (4), we get:

$$I_2 = 2\sum_{n=1}^{N}(n^2 + 4\alpha_n^2)\pi^2 A_{0n}^2; I_0 = \sum_{n=1}^{N} A_{0n}^2; \quad (5)$$

$$I_1 = 2\alpha_1 \sum_{n=1}^{N} \sum_{\substack{m=1 \\ m \neq n}}^{N} q_{nm} A_{0n} A_{0m} \sin^2(\pi(p_n - p_m)), \quad (6)$$

where $p_n = \alpha_n/\alpha_1$ ($p_1 = 1; p_2 = \alpha_2/\alpha_1 \equiv p; p_3 = \alpha_3/\alpha_1$ и т.д.). Convergence of the sum in (5) (n the expression for $I_2$) in the limit $N \to \infty$ is true under the following restriction on the initial amplitude: $A_{0n} \leq 1/n^{\frac{3+k}{2}}, k > 0$.

From (5), (6), we can get the following criterion of the linear instability of the PP flow:
$$\text{Re} > \text{Re}_{th} = I_2/I_1. \quad (7)$$

After minimization in the right-hand side of (7) over parameter $\alpha_1$, when the right-hand side in (7) reaches minimum for $\alpha_1 = \alpha_{1\min} = \sqrt{\frac{a}{b}}$, we get for the minimal threshold Reynolds number the following expression:

$$\text{Re}_{th\min} = \frac{\pi^4 \sqrt{ab}}{2c}, \quad (8)$$

$$a = \sum_{n=1}^{N} \frac{1}{n^{1+k}}; b = 4 \sum_{n=1}^{N} \frac{p_n^2}{n^{3+k}}; c = -\sum_{n=1}^{N} \sum_{\substack{m=1 \\ m \neq n}}^{N} (-1)^{n+m} \frac{\sin^2(\pi(p_n - p_m))}{(n^2 - m^2)^2 (nm)^{\frac{1+k}{2}}}. \quad (9)$$

The expression (8) shall be generally minimized over the free continuously changing parameters $k, p, p_3,..p_n..p_N$.

For simplicity, we restrict ourselves by the case when minimization in (8) is to done only over the first two parameters, $k$ and $p$, and the rest parameters we shall assume fixed by sufficiently slowly growing (for the convergence of the series (9) in the expression for $b$, when, for example, $p_n \cong n^{k/8}$ when $n=3,4, ...,N$).

2. Let us consider now (1) for the case of PC flow. It is important to note that in the representation of the solution of (1) in the form (2), the coefficients $B_n$ are non-zero. For example, in the case, $A_n = B_n$, we have

$$I_0 = 2\sum_{n=0}^{N} A_{0n}^2, I_2 = 2\pi^2 \sum_{n=0}^{N} A_{0n}^2 (4\alpha_0^2 p_n^2 + \frac{1}{2}(n^2 + \frac{(2n-1)^2}{4})), p_n = \frac{\alpha_n}{\alpha_0}, p_0 = 1, p_1 = p, p_n = n^{\frac{k}{8}}, n > 1$$

$$I_1 = 8\alpha_0 \sum_{n=0}^{N} \sum_{\substack{m=0 \\ m \neq n}}^{N} \frac{(-1)^{n+m} A_{0n} A_{0m} \sin^2(\pi(p_n - p_m))}{\pi^2} \left( \frac{n(m-\frac{1}{2})}{(n^2 - (m-\frac{1}{2})^2)^2} + \frac{m(n-\frac{1}{2})}{(m^2 - (n-\frac{1}{2})^2)^2} \right) \quad (10)$$

Using (10) and (7), we can find the stability threshold for the PC flow.

So, minimizing the right-hand side of (7) over parameter $\alpha_0$, we get again the formula (8), but now in it we already have ( e.g., when $A_{0n} = \frac{A_0}{(1+n)^{\frac{3+k}{2}}}$ )

$$a = \frac{1}{2} \sum_{n=0}^{N} \frac{(n^2 + \frac{(2n-1)^2}{4})}{(1+n)^{3+k}}, b = 4 \sum_{n=0}^{N} \frac{p_n^2}{(1+n)^{3+k}},$$

$$c = \sum_{n=0}^{N} \sum_{\substack{m=0 \\ m \neq n}}^{N} \frac{(-1)^{n+m}}{[(1+n)(1+m)]^{\frac{3+k}{2}}} \sin^2(p_n - p_m) \left[ \frac{n(m-\frac{1}{2})}{(n^2 - (m-\frac{1}{2})^2)^2} + \frac{m(n-\frac{1}{2})}{(m^2 - (n-\frac{1}{2})^2)^2} \right]. \quad (11)$$

Note that due to the consideration in (1) the disturbance velocity field transversal component only, the mass flux for the superposition of the main flow and the disturbance field is conserved. Natural emergence of such disturbance obviously is not excluded. In laboratory modeling of the PP and PC flow such disturbances may be easily created.

In the considered energy theory for amplitudes $A_n$, characterizing various transversal modes, we have used only restrictions related with the necessity of convergence of the sum $I_2$ in (5) and (10).

In the scaled form, fragments of the neutral curve corresponding to the condition (7)-(9) (see Fig. 1a) are given as dependences of the value $1/2p$ on $Re$ for the PP flow. On the Fig. 1b, the curves of neutral stability are given for the PC flow, where the criterion of linear (exponential) instability (7) was used taking into account (10) and (11).

## Comparison with the experimental data

For the PP, experimental results, e.g., [10], are known. In [10], contrary to the present work and [3], the Reynolds number threshold value is defined via the mean over the cross-cut flow velocity $V_a = \dfrac{2V_{max}}{3}$ and the full layer thickness $H$. In [10], the minimal threshold Reynolds number is found as $Re_D = \dfrac{V_a H}{\nu} = Re_{th\,exper} \approx 1440$. Thus, there is the following relationship: $Re_D = \dfrac{4Re}{3}$. Taking it into account, we get that the PP flow threshold value in [10] corresponds to the value $Re > Re_{th} \approx 1080$, that quantitatively differs less than by 4% from the estimate $Re_{th} \approx 1040$ obtained in the present paper. The best earlier known estimation with $Re_{th} \approx 5772$ was obtained by S.A. Orszag [2].

## Discussion and conclusions

The conclusion on the possibility of the linear instability of the PC and PP flows obtained in the present paper is inferred from the equation (1) for the evolution of the velocity disturbance field transversal component under condition that the right-hand side in (1) is non-zero due to the finiteness of the kinematic viscosity coefficient.

Earlier such a hydrodynamic dissipative instability mechanism was considered by L. Prandtl (1921-22) when studying stability of the laminar boundary layer and by V. Geisenberg (1924), and also by S.S. Lin (1944-45) when establishing the PP flow linear instability. In [8], an example of the linear two-dimensional oscillator instability (with linear on velocity friction) in a rotating reference frame is given that was considered as an example of secular instability [7]. In [8], additionally (in [7], attention was not brought to this) it was established relation of the pointed dissipative-centrifugal instability with the breaking of chiral symmetry defining the mechanism of the observed cyclone-anticyclone asymmetry in the atmospheres of the rotating fast planets. Understanding of the dissipative instability phenomenon may be also achieved with the help of L.D. Landau method (1941) he applied to assess the critical velocity of the superfluid motion in a capillary based on the consideration of the vortex disturbances (rotons) having negative energy.

Obtained in the present work and in [4-6] conclusions allow filling the known gap in the nonlinear theory [1], where instead of the seed linear exponential instability, only with necessity the stage of algebraic instability is considered.

Note also that it is interesting to consider application of the considered here linear instability of the PC flow to the problem of emergence of wave-killers in the ocean that are often observed in the regions with rather strong streams (Kurosio, Gulfstream, etc.) characterized by relatively large velocity shear [11]. And actually, earlier it was not considered possibility of the mechanism of such waves exciting related with hydrodynamic instability of the corresponding shear flows with respect to extremely small by amplitude disturbances. These velocity field disturbances according to the theory considered above may have a component transversal as to the horizontal flow velocity direction as well to the horizontal direction in which this velocity changes. It is clear that the velocity disturbance component $V_y$ for the considered in (1) PC flow can describe vertical fluid motions that reaching the water surface can cause emergence of anomalous by amplitude waves.

Although the known mechanism of the modulated instability [11-13] is based on the linear theory, it contains however the threshold condition on the disturbance amplitude for instability realization as the following inequality $A_0 > A_{0th} = \Delta\Omega/\omega_0 k_0 \sqrt{2}$ (in [11] this condition is written in another form as $\varepsilon N > 1/\sqrt{2}, \varepsilon = k_0 A_0, N = \omega_0/\Delta\Omega$), where $A_0, k_0, \omega_0, \Delta\Omega$ are initial amplitude, wave number, wave disturbance frequency and modulation frequency respectively. It is important that in the considered in the present paper mechanism related with the FC flow linear instability, any conditions on the disturbance amplitude are absent.

It is interesting to consider in the future application of the developed above energy method to the assessment of the minimal threshold Reynolds number for the cylinder Taylor-Couette flow that has not the problem of linear stability known for PC and HP flows. Alos, it is desirable also conducting of direct numerical computations of the PC and PP flows stability similar to those conducted in [14, 15], aiming establishing applicability range of the energy method.

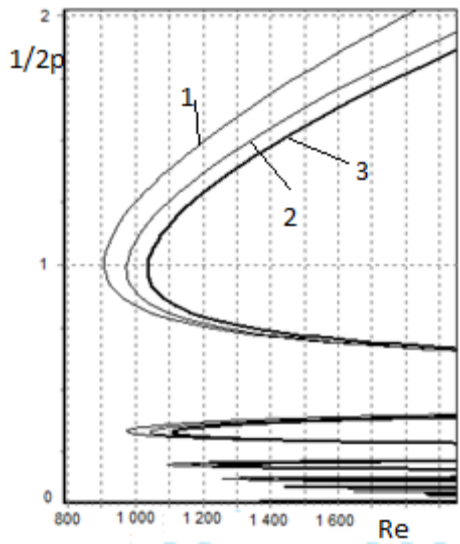 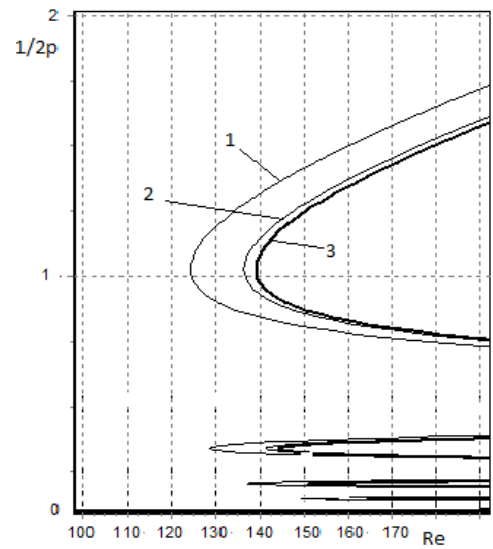

a) b)

Fig. 1.

For the PP flow on Fig. 1a), curves of the neutral stability re given for $k=0.675$. For $N=2$ (curve 1, minimum is *906.35, 1/2p=1.008*), $N=10$ (curve 2, minimum is *972.825, 1/2p=0.988*), $N=100$ (curve 3, minimum is *1035,311, 1/2p=0.988*).

For the PC flow on Fig. 1b), the curves of the neutral stability are given for $k=1.7037$, N=2 (curve 1, minimum is *124.273, 1/2p=1.029*), $N=10$ (curve 2, minimum is 136.475, 1/2p=1.029), $N=100$ (curve 3, minimum is *139.077, 1/2p=1.029*).